\documentclass[11pt]{article}
\usepackage{amsmath,amssymb}
\usepackage{hyperref}
\newsavebox{\uuunit}
\sbox{\uuunit}
    {\setlength{\unitlength}{0.825em}
     \begin{picture}(0.6,0.7)
        \thinlines
        \put(0,0){\line(1,0){0.5}}
        \put(0.15,0){\line(0,1){0.7}}
        \put(0.35,0){\line(0,1){0.8}}
       \multiput(0.3,0.8)(-0.04,-0.02){10}{\rule{0.5pt}{0.5pt}}
     \end {picture}}

\def\bea{\begin{eqnarray}}
\def\eea{\end{eqnarray}}

\numberwithin{equation}{section}

\begin{document}

\thispagestyle{empty}
{}

\vskip -3mm
\begin{center}
{\bf\LARGE
\vskip - 1cm
Boundary Weyl anomaly of  $\mathcal{N}=(2,2)$ superconformal models \\ [2mm]}

\vspace{10mm}

{\large
{\bf Constantin Bachas} \  and\,\    
{\bf Daniel Plencner} 
 
\vspace{1cm}

{ 
Laboratoire de Physique Th\'eorique de l' \'Ecole Normale Sup\'erieure,\\ 
 PSL Research University,  CNRS,  Sorbonne Universit\'es, UPMC Univ.\,Paris 06,\\ 
24 rue Lhomond, 75231 Paris Cedex 05, France \\
\vskip 4mm
}}

\vspace{5mm}

\end{center}
\vspace{-2mm}

\centerline{\it Dedicated to the memory of Ioannis Bakas}

\vspace{8mm}

\begin{center}
{\bf ABSTRACT}\\
\end{center}

 We calculate the trace and axial anomalies  of    $\mathcal{N}=(2,2)$ superconformal theories
 with exactly marginal deformations, 
 on a surface with boundary.  Extending recent work by Gomis et al, we derive  the boundary contribution 
 that captures   the  anomalous scale dependence  of the one-point functions of exactly marginal operators.
 Integration of the bulk super-Weyl anomaly   shows that  the sphere partition function computes the K\"ahler potential 
  $K(\lambda, \bar\lambda)$   on  the superconformal manifold.  
  Likewise,  our results confirm the conjecture  that the  partition function 
 on the  supersymmetric hemisphere computes the
  holomorphic  central charge,  $c^\Omega(\lambda)$,  associated with the boundary condition $\Omega$. 
 The boundary entropy, given by a ratio of hemispheres and sphere, is therefore  fully determined    by  anomalies.

\clearpage
\setcounter{page}{1}


\section{Introduction and Summary}

Recently  Gomis et al  \cite{Gomis:2015yaa}  computed the trace anomaly of two-dimensional
${\cal N}=(2,2)$  superconformal field theories (SCFT) that belong to   a continuous family
 (alias  `superconformal manifold')  ${\cal M}$. 
The complex    moduli  $\{ \lambda^I\}$ that parametrize  ${\cal M}$
 couple to exactly marginal    operators $\{ {\cal O}_I\}$ whose 
   2-point functions  have a scale anomaly   first discussed  by 
Osborn  \cite{Osborn:1991gm}. This  anomaly only manifests itself when the  $\lambda^I$  vary in   space,
 but   supersymmetry  relates it   to another  term 
 that does not vanish   for constant couplings, 
   $  \delta \int  \sqrt{g} \, R^{(2)}  K(\lambda, \bar\lambda)$ where $K$ is the K\"ahler potential on ${\cal M}$. 
 Although  this term 
  is the variation of a local action,   it cannot be  supersymmetrized in a local way
 and is hence an irreducible part  of the anomaly. The   integrated anomaly thus  contains a
 term ${1\over 2} K \chi_M$ with $\chi_M$  the Euler characteristic of $M$. 
 
 An important corollary is that  the  sphere partition function has a universal finite piece  
  that does not depend on the renormalization scheme.\footnote{Actually there exists
   a residual  ambiguity \cite{Gomis:2015yaa}  that  corresponds to K\"ahler-Weyl transformations of $K(\lambda, \bar\lambda)$. 
   This will be important in our discussion later on.}
  Explicitly
        \bea\label{01}
  {\cal Z}(S^2) =   \left({r\over r_0}\right)^{c/3} e^{-K(\lambda, \bar\lambda)}\  ,
  \eea
where $r$ is the radius of the sphere, $r_0$ a short-distance cutoff and $c$ the central charge of the theory.
In many cases ${\cal Z}(S^2)$ can be  computed exactly using  localization  of the path integral
\cite{Benini:2012ui, Doroud:2012xw},  which is thus established as  a powerful new tool for the
calculation of  worldsheet-instanton corrections to $K$,  and of  the associated Gromov-Witten invariants. 

\smallskip

 The relation  \eqref{01} was first conjectured by  Jockers et al \cite{Jockers:2012dk} and established
 by different methods in
\cite{Gomis:2012wy, Gerchkovitz:2014gta}. The proof  based on the trace anomaly is elegant and more powerful, 
  it can be generalized for example  to four dimensions \cite{Gomis:2015yaa}. 
In this  paper we will extend the calculation of the anomaly  in another direction, allowing for  a surface  with boundary. 
As a corollary we will derive  the  hemisphere relations
        \bea\label{02}
  {\cal Z}_+(D^2) =   \left({r\over r_0}\right)^{c/6}   c^\Omega(\lambda)\  ,
  \qquad
   {\cal Z}_-(D^2) =   \left({r\over r_0}\right)^{c/6}   c^\Omega(\bar\lambda)\  ,
  \eea
where  $\pm$ indicate a choice of spin structure,  and $c^\Omega(\lambda)$ is the holomorphic 
central charge of the boundary condition $\Omega$. 
These  relations have been conjectured  by Honda and  Okuda
  \cite{Honda:2013uca} and  by Hori and Romo \cite{Hori:2013ika}
 (see also ref.\,\cite{Sugishita:2013jca, Fujimori:2015zaa}).  Here we will show that they follow   from
  the supersymmetric trace anomaly.

A key ingredient in our calculation is the scale anomaly of the 
 one-point functions  $\langle {\cal O}_I\rangle_\Omega$,  whose 
  supersymmetrization gives a term  proportional to the boundary charge $c^\Omega$. 
   Without supersymmetry this term would have been ambiguous, much like the term proportional to $K$
  in the bulk. Thanks to supersymmetry it acquires universal  meaning. 
   The  vacuum degeneracy of the boundary  (alias $g$-function \cite{Affleck:1991tk})   is   
$
g^\Omega = \vert c^\Omega \vert\, e^{K/2}\, . 
$
 Our result therefore shows  that  
 \bea\label{03}
 g^\Omega  =   \sqrt{ {\cal Z}_+(D^2) {\cal Z}_-(D^2)/{\cal Z}(S^2)} \ . 
 \eea
 As a special application, one can
  use localization to compute Calabi's diastasis function \cite{Bachas:2013nxa}.

       We now describe these results in   more detail. The  $\mathcal{N}=(2,2)$ superconformal theories  
 are defined on a Riemann surface  $M$ with boundary.  
 Boundary conditions, $\Omega$, that preserve  half  of the superconformal symmetry are of  two types,
 A and  B. For  SCFTs  whose target space   is a large Calabi-Yau threefold  
 they correspond,  respectively, 
 to special-Lagrangian and to
 holomorphic submanifolds   \cite{Ooguri:1996ck} (see also 
   \cite{Hori:2000ck, Lindstrom:2002jb}). Mirror symmetry exchanges A with B,   so we may   
restrict  to  the B-type  boundaries only.

The moduli space   of $\mathcal{N}=(2,2)$  SCFTs  factorizes locally\footnote{Except when the supersymmetry is enhanced,  as discussed in the recent reference \cite{Gomis:2016sab}. }
   into chiral and twisted-chiral  deformations, ${\cal M} \simeq {\cal M}_{\rm c}\times {\cal M}_{\rm tc}$. 
  In the geometric Calabi-Yau limit
   these factors correspond, respectively,  to    complex-structure and 
K\"ahler moduli. Chiral deformations are generically obstructed at a B-type boundary, i.e.~supersymmetry is
completely broken  \cite{Fredenhagen:2006dn, Baumgartl:2007an}. 
 There is no such obstruction for twisted chiral deformations but another, non-generic 
phenomenon can occur for them: across  lines of marginal stability  the bulk deformation can  induce a boundary renormalization-group flow
\cite{Brunner:2009mn}, so that physics at the boundary is 
discontinuous.\footnote{Such a  flow is induced
whenever the marginal bulk operator mixes with relevant or marginal  operators on the 
boundary \cite{Fredenhagen:2006dn}.
 In string theory 
these  flows are interpreted as  D-brane decays.}   
 To avoid these complications  we will restrict attention  to the (unobstructed)  twisted chiral moduli only, 
 and 
 to   regions of
their moduli space that are  free from 
marginal-stability lines. The $\lambda^I$    in \eqref{02}  
are  the  twisted-chiral couplings inside such  regions.

   In   calculating  the boundary anomaly we rely on   different  arguments. 
 These include  parity invariance, 
  the covariance of the standard trace anomaly,  and supersymmetric Ward identities.
  An  important consistency check
  is that different  K\"ahler-Weyl frames  should correspond to different
 renormalization schemes, i.e.~that they should be  related by the addition of local gauge invariant counterterms \cite{Gomis:2015yaa}.
  Another 
  non-trivial check   is that the one-point functions of marginal operators, 
  $\langle {\cal O}_I\rangle_\Omega$, 
        have the same dependence on  $K$ and $c^\Omega$  as 
 the  Ramond-Ramond charges
  of Calabi-Yau D-branes  \cite{Ooguri:1996ck, Hori:2000ck}. 
 We will see why this is no   coincidence.
  
   \smallskip
       
   The  paper is organized as follows. In section 2 we review  the calculation of  ref.\,\cite{Gomis:2015yaa}
   paying  particular attention to the scale  anomaly of  two-point functions,  and to the role of K\"ahler-Weyl
   transformations. Section 3 introduces  two simple tricks   for restoring  supersymmetry in the presence of boundaries: 
   A standard `reference'  completion of  D-term integrals, and  
   boundary superspace \cite{Hori:2000ic, mirror}.
    Section 4  presents  the general  ansatz for the boundary anomaly that is   parity invariant,  and 
  consistent with a special supersymmetric Ward identity that we will explain.   The ansatz  depends on a holomorphic function of the moduli which we identity with the 
  logarithm of the boundary  charge $c^\Omega$. This identification is confirmed in
     section 5 where we extract  the  moduli-dependence
     of the one-point functions on the half-plane, and show that they  agree with the Ramond-Ramond charges
     of $\vert \Omega \rangle \hskip -0.5mm \rangle$ derived in  \cite{Ooguri:1996ck}. 
     Asking that 
              K\"ahler-Weyl transformations   be
     cohomologically trivial implies that $c^\Omega$ is  the section of a holomorphic line bundle.
      In section 6 we integrate the anomaly for the supersymmetric squashed hemisphere \cite{Gomis:2012wy}
     and establish the relations   \eqref{02}. 
      The result does not depend on the squashing parameter.
        In  appendix A  we summarize 
        conventions.


\section{Review of the Bulk Anomaly}
 
 We begin  by reviewing the results of Gomis et al
 \cite{Gomis:2015yaa}. We consider the $U(1)_V$ supergravity whose
gauge field $V^\mu$ couples to the vector-like R-symmetry, the one preserved by the
 B-type boundary conditions. The theory is defined on a surface $M$ (not to be confused with the moduli space
 ${\cal M}$). 
For  details  of the $\mathcal{N}=2$, $d=2$ supergravity see  \cite{Closset:2014pda}.
 In superconformal gauge the graviton multiplet has two real bosonic
degrees of freedom, $\sigma$ and $a$,  which are defined by 
$g_{\mu\nu} = e^{2\sigma}\eta_{\mu\nu}$ and 
$V^\mu = \epsilon^{\mu\nu}\partial_\nu a$. They combine to form the lowest 
component of a twisted chiral field\footnote{Our superfield conventions
are given in  appendix A. Note that in ref.\,\cite{Gomis:2015yaa} twisted chiral fields 
and twisted chiral  couplings are denoted by tildes in order to distinguish them from  chiral fields and chiral couplings. 
Since in this paper we  focus on the twisted-chiral sector only, the tildes are dropped. 
}
\bea
\Sigma(y^\mu)  = (\sigma + i a)  +  \theta^+ \bar\chi_+ + \bar\theta^- \chi_- + \theta^+  \bar \theta^- w \ , 
\eea
 where component fields are  
 functions of the coordinates $y^\pm = x^\pm \mp i \theta^\pm \bar\theta^\pm$. 

This  $U(1)_V$ supergravity is coupled to a ${\cal N}=(2,2)$ superconformal field theory which has a set of marginal 
  couplings $\lambda^I$ that parametrize the superconformal manifold. As explained in the introduction, we
  restrict ourselves to the twisted chiral deformations. A useful trick \cite{Seiberg} is to 
   promote  the $\lambda^I$
   to  expectation values of 
   the  (lowest components
of)  twisted chiral superfields  $\Lambda^I$. 
We will be  interested in the anomalous dependence of the partition function ${\cal Z}_V(M)$ on the  Weyl 
superfield $\Sigma$. 
 The two  supersymmetries and the vector-like R-symmetry can be preserved by the regulator,  so
the anomaly, \,  $ i A({\delta\Sigma}) :=   \delta_{\Sigma} \log {\cal Z}_V(M)$,   must be  
 an  invariant  combination  of the twisted-chiral  superfields $\Sigma$ and $\Lambda^I$. 
\smallskip
 
 For   a closed Riemann surface the
   anomaly has been computed   in 
   ref.\,\cite{Gomis:2015yaa}. It is  the sum of two separately-invariant  terms,\footnote{This 
   expression  differs  from the one given in \cite{Gomis:2015yaa}  by a pure twisted-chiral  term and its
   complex conjugate, \, 
   $ 
     \int_{ M} \hskip -1mm d^2x  \hskip -0.1mm  \int \hskip -1mm  d^4\theta \,
     (\delta   \Sigma\,  { \Sigma} +   \delta  \bar{  \Sigma}\, \bar \Sigma )\, ,   
   $ \ 
which  integrate to zero when $M$ is  closed. The overall sign  corresponds  to the case of twisted chiral fields
(the ones with tildes) 
in this reference.
}
   \bea\label{1}
A_{\rm closed}   = {1\over 4\pi} \int_{ M} \hskip -1mm d^2x  \hskip -1mm  \int \hskip -1mm  d^4\theta \,
\left[     \frac{c}{6} (\delta   \Sigma\, \bar{ \Sigma} +   \delta  \bar{  \Sigma}\,\Sigma )
- 
 (\delta   \Sigma+\delta  \bar{  \Sigma}) 
  K( \Lambda,\bar{ \Lambda}) \right]    := A^{(1)}   + A^{(2)} \ , 
\eea
where $A^{(1)} $ is proportional to the central charge
 $c$ of the SCFT,  while $A^{(2)} $  depends on the marginal couplings
 via the K\"ahler potential  $K( \lambda,\bar{ \lambda})$ of 
the superconformal manifold. 
 The  anomaly  satisfies  the
   Wess-Zumino consistency condition   
$ \delta_\Sigma A({\delta\Sigma^\prime}) - \delta_{\Sigma^\prime} A({\delta\Sigma})  = 0$,  
and  can thus be   integrated with the result
\bea\label{Integr}
  \log {\cal Z}_V  \  \supset  \  \ {i\over 4\pi}  \int_{ M} \hskip -1mm d^2x  \hskip -1mm  \int \hskip -1mm  d^4\theta \,  
 \left[ {c\over 6} \Sigma\, \bar{ \Sigma}\, -   \, (  \Sigma+  \bar{  \Sigma}) 
  K \right]\ .  
\eea
Though local in superconformal gauge,   the right hand side
 cannot  be written  as the integral of a  local {\it covariant}  density -- it is  cohomologically non-trivial. 
 The  term  proportional to $c$   is the canonical  
 kinetic action for the  ${\cal N}=2$  Liouville
superfield $\Sigma$, while the  less familiar second term is determined  by 
 comparison with the two-point function 
of marginal operators on the plane,  as we now explain.
\smallskip


\subsection{Anomalous two-point functions and ${\cal Z}(S^2)$}

  The  meaning of the anomaly \eqref{1}  is more transparent   in component form. 
  We have defined  $A^{(1,2)}$ as the D-term superspace  integrals, 
   $$
   A^{(1)}:=  {c\over 24\pi} \int_{ M} \hskip -1mm d^2x  \hskip -1mm  \int \hskip -1mm  d^4\theta \,
  (\delta   \Sigma\, \bar{ \Sigma} +   \delta  \bar{  \Sigma}\,\Sigma )\quad   
  {\rm and}  \quad A^{(2)}:=  -{1\over 4\pi} \int_{ M} \hskip -1mm d^2x  \hskip -1mm  \int \hskip -1mm  d^4\theta \,
  (\delta   \Sigma +   \delta  \bar{  \Sigma}  ) K(\Lambda, \bar\Lambda)\ .  
  $$ 
Expanding   $K(\Lambda^I(y^\pm), \bar\Lambda^I(\bar y^\pm))$
  in Grassman variables, and  
setting  the fermionic and auxiliary components of  the  coupling superfields  to 
zero gives
 \bea\nonumber
A^{(1)} 
 = - \frac{c}{12\pi} \int_M d^2 x\,  \Bigl[\,   \delta \sigma\,  \square \sigma +  \delta a\,  \square a
 + \frac{1}{2}(\delta w \,\bar w + \delta\bar w \, w) + \partial^\mu b^{(1)}_\mu  
\Bigr]\ + {\rm fermions}\  ,   
 \eea 
\bea\nonumber
A^{(2)} 
 = - \frac{1}{2\pi} \int_M d^2 x\,  \Bigl[ \delta\sigma\, (\partial_\mu \lambda^I \partial^\mu \bar \lambda^{\bar J}) 
  \partial_{I }\partial_{ \bar J}K \,  - \frac{1}{2} K\, \square \delta \sigma - (\partial^\mu \delta a )   {\mathcal{K}}_\mu + \partial^\mu b^{(2)}_\mu
   \Bigr]\  
\eea
\vskip -5mm
\bea\label{2}
{\rm where}\qquad   {\mathcal{K}}_\mu := {i\over 2} (\partial_I K \partial_\mu \lambda^I  - \partial_{\bar I}  K \partial_\mu \bar \lambda^{\bar I})\  .  
\eea
The total-derivative terms, which integrate to zero when $M$ is closed, are given by
\bea\nonumber
b^{(1)}_\mu =   {1\over 4} (\partial_\mu \delta\sigma) \sigma -{3\over 4} \delta \sigma\,   \partial_\mu\sigma +
{1\over 4} (\partial_\mu \delta a)  a -{3\over 4} \delta  a\,  \partial_\mu a\ , 
\eea \vskip -5mm
\bea\label{5} 
b^{(2)}_\mu =   {1\over 4} (\partial_\mu \delta\sigma) K  -  {1\over 4}   \delta\sigma\, \partial_\mu  K \ .   
\eea
We keep  them because they  will be needed  for  surfaces with boundary.\footnote{Note that
 the Wess-Zumino consistency --
  manifest in  the superfield expression \eqref{1}, would be violated
if these  terms were dropped. For example, 
 the contribution to   the pure Weyl anomaly  
$ 
 \int \delta\sigma \square \sigma + \partial^\mu b^{(1)}_\mu = {1\over 4} \int \left[
    \delta\sigma \square \sigma
+  \sigma \square \delta \sigma   - 2(\partial_\mu \delta\sigma) (\partial^\mu \sigma) \right]
$\,,   is symmetric  under   $\sigma\leftrightarrow \delta\sigma$, as it should,   
only if  these boundary   terms are included.}

\smallskip

The first  line in  \eqref{2} gives  the well-known Weyl anomaly\, $  - {c\over 12\pi}  \int  \delta \sigma\,  \square \sigma =
{c\over 24\pi} \int  \delta \sigma\,  \sqrt{g} R^{(2)}    $,   
and its  ${\cal N}=2$ supersymmetric completion. The less familiar second line
begins  with a  term 
that vanishes when the  $\lambda^I$ are constant. 
  This anomaly, first discussed in ref.\,\cite{Osborn:1991gm} by Osborn, 
   captures the logarithmic divergence of the  
   {   regularized two-point functions of the   marginal operators
  ${\cal O}_I$. More explicitly one has  
\bea\label{2point}
\langle {\cal O}_I(z) \bar {\cal O}_{\bar J}(w)\rangle\  = \   g_{I\bar J} \, {\cal R} {1 \over \vert z-w\vert^4}\
= \ g_{I\bar J} \,   {1\over 2} (\partial  \bar \partial )^2 \bigl[\log(\vert z-w\vert^2 \mu^2) \bigr]^2\ , 
\eea   
where  $g_{I\bar J}  =  \partial_I \partial_{\bar J} K $ is the Zamolodchikov metric  \cite{Zamolodchikov:1986gt}
at the unperturbed point,
$\lambda^I=0$, of the superconformal manifold, and the symbol ${\cal R}$ denotes   the differential
regularization of the two-point function \cite{Osborn:1991gm} 
which should be  viewed as a distribution.\footnote{The reader
can easily check that  at $z\not= w$  the regularized two-point function equals $\vert z-w\vert^{-4}$. 
By pulling   the derivatives in the front  
 one ensures that they can be transferred to test functions of $z$ or $w$. If these test functions are 
   sufficiently smooth,   the singularity
at $z=w$ is integrable.
 } In  the normalization  of \cite{Gomis:2015yaa} one finds for the Euclidean  generating  function
 \bea
 -\mu {\partial \over \partial \mu}\log {\cal Z}_V^E  \  \supset  \ 
 -\mu {\partial \over \partial \mu} \int {d^2z\over \pi}\int {d^2w\over \pi}\, \lambda^I(z,\bar z) \, \bar \lambda^I(w,\bar w)
 \   g_{I\bar J} \, {\cal R} {1 \over \vert z-w\vert^4}\ . 
\eea     
 Using the identities
  $\square =  4 \partial  \bar \partial$ and $ \partial  \bar \partial  \log\vert z\vert^2 = \pi \delta^{(2)}(z)$ it can be checked
 that this 
  agrees precisely with   $A^{(2)}$ for   $\delta\sigma = -\delta\log\mu$\ \,    constant. 
  }  
   
 \smallskip
    
     This   argument allows us to identify  the
 (a priori arbitrary)  function $K(\lambda, \bar\lambda)$   with
 the K\"ahler potential 
 of  the superconformal manifold. 
    Now the anomaly $A^{(2)} $ also contains   the term 
    $\int {1\over 4\pi}  K\, \square \delta \sigma$,    which does not vanish when the $ \lambda^I $ are constant. 
     Taken in isolation  this  term  would have   been cohomologically  trivial since it is 
 the variation of
        $ -{1\over 8\pi}  \int_M  \sqrt{g} R^{(2)}\, K $,   where $R^{(2)}$ is the Ricci scalar of the Riemann surface. 
            Supersymmetry relates it however  to the cohomologically non-trivial term 
       $\sim \delta\sigma \vert \partial \lambda\vert^2$,  
      so that ${\cal N}=2$ invariant counterterms  cannot  remove it.  
      This is the key observation  in ref.\,\cite{Gomis:2015yaa}.
One important corollary is  that the free energy on the round two-sphere gives the 
  K\"ahler potential  of the conformal manifold, 
\bea\label{sphere}
{\cal Z}_V^E(S^2) =   \left({r\over r_0}\right)^{c/3} e^{-K(\lambda, \bar\lambda)}\ , 
\eea
 where $r$ is the radius of the sphere,  $r_0$ is an ultraviolet cutoff and we used the identity  
 $\int_{S^2} K\, \square   \sigma = -4 \pi K$ ,   valid when $K(\lambda, \bar\lambda)$ is constant
 on $S^2$. 
  This 
 relation  
 was first conjectured by Jockers {  et al}   \cite{Jockers:2012dk},
  and proved by a different argument in \cite{Gomis:2012wy, Gerchkovitz:2014gta}. 
   The above proof based on the anomaly
  is   powerful and appealing.

  The importance of eq.\,\eqref{sphere} stems from the fact that  the sphere partition function  
 can be sometimes obtained  exactly by  localization
 of the functional integral  \cite{Benini:2012ui, Doroud:2012xw} (as pioneered  in
 \cite{Pestun:2007rz}, 
 see  also  \cite{Benini:2016qnm, Park:2016dpb}
 for recent reviews). 
  This is  a powerful new  method for computing the quantum K\"ahler potential  on the moduli
  space of Calabi-Yau threefolds, 
   which does not  rely  on mirror symmetry.


\subsection{K\"ahler-Weyl  transformations}

    An  important remark about the above formula   concerns the  
K\"ahler-Weyl  transformations 
\bea\label{KW}
K^\prime (\lambda, \bar \lambda)\, = \,  K(\lambda, \bar \lambda) + H(\lambda) + \bar H(\bar \lambda) \ ,   
\eea
where   $H$  is a  holomorphic function of the $\lambda^I$. Such transformations do not affect the geometry of
 the superconformal manifold but they do modify  the anomaly \eqref{1}, 
 \bea\label{KW1}
 \Delta_{\rm KW} A^{(2)} = - {1\over 4\pi} \int_{ M} \hskip -1mm d^2x  \hskip -1mm  \int \hskip -1mm  d^4\theta \, (\delta   \Sigma  +\delta  \bar{  \Sigma}) H  + c.c. \ . 
\eea
This looks paradoxical, at first sight, since physics should be independent of the K\"ahler-Weyl frame. 
The puzzle is elegantly resolved \cite{Gerchkovitz:2014gta, Gomis:2015yaa} 
by  noting that the right-hand side  can be  equivalently
  rewritten as a (twisted)  F-term,  
 $$
 \int_{ M} \hskip -1mm d^2x  \hskip -1mm   \int \hskip -1mm  d^4\theta \, (\delta   \Sigma  +\delta  \bar{  \Sigma}) H  =
 \int_{ M} \hskip -1mm d^2x  \hskip -1mm  \int \hskip -1mm  d \theta^+ d\bar \theta^-  \, \bar D_+D_- [(\delta   \Sigma  +\delta  \bar{  \Sigma}) H] \, + \, \int_{ M} \hskip -1mm d^2x \  (\partial^\mu Y_\mu)
 $$ \vskip -5mm
 \bea
  =
\int_{ M} \hskip -1mm d^2x  \hskip -1mm  \int \hskip -1mm  d \theta^+ d\bar \theta^-  \,  (\bar D_+D_- \delta  \bar{  \Sigma}) H +  \, \int_{ M} \hskip -1mm d^2x \  (\partial^\mu Y_\mu)\ ,  
 \label{nobnrs}
 \eea
where  $\partial^\mu Y_\mu$ is a total divergence whose integral over a closed surface   $M$ vanishes. 
We used here the identity $ \int \hskip -1mm  d^4\theta \, X  = \int \hskip -1mm  d \theta^+ d\bar \theta^- 
 \, \bar D_+D_- X  + \partial^\mu Y_\mu $ , valid for any superfield $X$, as well as  the fact that  $\bar D_+$ and $D_- $ 
 annihilate the twisted chiral superfields   $\delta\Sigma$ and $H(\Lambda)$. 
 
   Now
 $
 \bar D_+D_-   \bar \Sigma$   is a  twisted-chiral  superfield  whose components are
    the  curvature and the 
 field strength of the  $U(1)_V$ gauge field,  
\bea\label{Rsuperfield}
\bar D_+D_-  \bar \Sigma\, =\, -\bar w\,  +  \,  4 \theta^+ \bar\theta^-  {\partial_+\partial_-} (\sigma -  ia)  + \cdots\ \  .  
\eea
Thus  \ $\int_{ M} d^2x \hskip -0.7 mm  \int \hskip -1mm d \theta^+ d\bar \theta^-\,  \bar D_+D_-   \bar \Sigma\, H(\Lambda)
   + c.c.\ $\  is a 
 local,  supersymmetric,  reparametrization-invariant counterterm \cite{Gerchkovitz:2014gta}  
 which can be used to cancel 
  the K\"ahler-Weyl transformation  \eqref{KW1}. Let us stress this again:
   the renormalization-scheme  ambiguity  
 translates into a   
   freedom of  K\"ahler-Weyl  transformations  which are thus cohomologically trivial. 

It is   instructive to also verify this   in component form. The K\"ahler-Weyl transformation changes
${\cal K}_\mu$ by the total derivative ${i\over 2} \partial_\mu (H - \bar H)$. 
The change in the anomaly \eqref{2} is therefore, after integration by parts, proportional to 
$ (H + \bar H) \square \delta\sigma
+   i (\bar H -H) \square \delta a$. Since $\square a = -\epsilon^{\mu\nu} \partial_\mu V_\nu$ 
 and $2 \square \sigma = - R^{(2)}\sqrt{g}$, both terms can be cancelled by local gauge- and reparametrization-invariant    counterterms,  as advertized.  

The fact that  a K\"ahler-Weyl transformation amounts   to a change of regularization scheme leads to
an  interesting conjecture \cite{Gomis:2015yaa}. If   the superconformal manifold 
 had non-vanishing K\"ahler class, one could 
choose space-dependent couplings $\lambda^I(x)$
such that  the embedding of $M \subset {\cal M}$  is  a  nontrivial
2-cycle. It would then be impossible to find a regularization scheme, valid everywhere
in $M$, for such  sigma models.   
To avoid  this embarrassing  situation  one must demand that
 ${\cal M}$  have  
vanishing   K\"ahler class, a   non-trivial restriction on 
superconformal manifolds.


 \section{Supersymmetry with  Boundaries}
 
    Having summarized the anomaly in the bulk, we turn now our attention 
     to the case when $M$ is an open Riemann surface. 
           The presence of the boundary affects  the calculation   in different  ways. 
          Firstly,   the component expressions \eqref{2} include 
total-derivative terms which   cannot be ignored 
 when   $M$ is  open. Secondly, these expressions are  not invariant under supersymmetry
 anymore,  because the top component of a superfield
  transforms to a total derivative under a supersymmetry transformation. To cancel this  we  must  add
  a compensating boundary term, as 
   is well-known from   the study of ${\cal N}=2$ sigma-model actions (see for example 
  \cite{Brunner:2003dc}). 
  Of course, supersymmetry alone does not suffice to fix 
 the boundary  anomaly  uniquely. There exist several  candidates that differ by 
  supersymmetric boundary  invariants,  and our   task will be to find the right one.

  We will proceed in two steps. In this section we  describe a general algorithm
  that  restores the invariance  of a D-term   superspace  integral  $\int_M d^2x \int d^4\theta {\cal S}$, 
  for any superfield ${\cal S}$, when $M$ has a boundary. 
  All  other  completions  differ  from this  `reference'  one by a superinvariant which can be always 
  written as  an integral over boundary superspace  \cite{Hori:2000ic, mirror}.  To find the right one we  
  will need extra  arguments that  are  presented  in sections 4 and 5.


 \subsection{Reference completion of bulk D-terms}

 { 
 Let us consider the  integral 
 $ \int_M \hskip -0.1mm d^2x \int \hskip -1mm d^4\theta \,    {\cal S}  =   \int_M \hskip -0.1mm d^2x\,  [{\cal S}]_{\rm top}  $, 
 where ${\cal S}$ is a  real  superfield and  $[{\cal S}]_{\rm top}$ is  its $\theta^4$  component. 
  The B-type supersymmetry transformations, the ones
 consistent with B-type boundary conditions,  are 
 generated by  the differential operators in superspace
 \bea\label{Dsusy}
{\cal D}_{\rm susy} =  \epsilon\, ( e^{i\beta} Q_++ e^{-i\beta}Q_-) - \bar\epsilon\, (e^{-i\beta}\bar Q_++ e^{i\beta}\bar Q_-)\ , 
\eea
where   $\epsilon$ is a complex  Grassmanian parameter, $\beta$ is an arbitrary phase,  and 
 \vskip -6mm  
\bea
{\cal Q}_\pm = {\partial \over \partial \theta^\pm} + i \bar\theta^\pm \partial_\pm\ ,
\qquad
\overline{\cal Q}_\pm = - {\partial \over \partial \bar \theta^\pm} - i  \theta^\pm \partial_\pm\   
\eea
(see appendix A). Note that  the   vector and axial $R$-symmetry  charges  are $(1, 1)$ for $\bar Q_+$, and $(1, -1)$
for $\bar Q_-$\,.  If $M$ has  a single boundary, one  can always reabsorb $\beta$ by a redefinition
of the fermions.  We   set here $\beta=0$. We can always restore it later,  if needed. 
}

 The transformation of $[{\cal S}]_{\rm top}$ is a total derivative,  
\bea\label{Stop}
{\Delta}_{\rm susy} [{\cal S}]_{\rm top}  =   \int \hskip -1mm d^4\theta \, \, {\cal D}_{\rm susy} {\cal S}
=  \, i \epsilon  \int \hskip -1mm  d^4\theta \,\,  
  ( \bar\theta^+ \partial_+{\cal S}  +  \bar\theta^- \partial_-{\cal S} )\  +\  c.c.\ , 
\eea
 which integrates to zero
when  $M$  is closed,  but need not vanish  when $M$ is open. 
Let's assume that $M$ is   the
 half-space $x^1\leq 0$. We 
  would like to express the right-hand side of \eqref{Stop} 
as the   transformation of a spatial  derivative, up to  a time derivative,  
$${\Delta}_{\rm susy} [{\cal S}]_{\rm top} =
-{\Delta}_{\rm susy} (\partial_1 [{S}]_{\rm bnry}) + \partial_0 Y \ , 
$$ 
  so that  \vskip -8mm
\bea\label{I}
    {\rm I}_D({\cal S}) :=  \int  \hskip -0.1mm d^2x\,    [{\cal S}]_{\rm top} +  \int dx^0 \,[{S}]_{\rm bnry} 
\eea   is  a supersymmetric  invariant. It  will serve as  our reference completion.

\smallskip

 To  bring   \eqref{Stop}  to the above form  we proceed as follows. Begin with the identity   
\bea\label{7}
  \int \hskip -1mm d^4 \theta \,\,  Q_+ {\cal S} =  \int \hskip -1mm d^4 \theta \, \, i   \bar \theta^+ \partial_+ {\cal S} = 
  \int \hskip -1mm d^4\theta \,   i  
   \bar\theta^+    \theta^-   ( Q_-  - i \bar\theta^- \partial_-) 
   \partial_+  {\cal S} \ , 
\eea
where the second equality follows from the rule of  integration by parts in Grassman space,
 $ \int d^4 \theta \, \bar\theta^+   \theta^- {\partial X/ \partial \theta^-} = \int d^4 \theta \, \bar\theta^+ X$\ 
 for any $X$.  Adding the vanishing term $ \int d^4\theta \,   i  \bar\theta^+   \theta^-  Q_+ \partial_+ {\cal S}$,
 and inserting $\theta^+ Q_+$ in the last term (this  just acts as the identity within the superspace integral) 
 allows us to rewrite this equation as 
 $$
 \int \hskip -1mm d^4 \theta \,\,   Q_+{\cal S} = \int \hskip -1mm d^4\theta \,\,   i  \bar\theta^+   \theta^-  ( Q_+ + Q_-) \partial_+  {\cal S} - \int \hskip -1mm d^4\theta \,\, \theta^4   Q_+ \partial_- \partial_+ {\cal S},
$$
where  $\theta^4 = \theta^+ \theta^- \bar \theta^- \bar\theta^+$. 
 This  identity also holds if we exchange $+$ with $-$ indices, so that
 \begin{align}  \label{eq:Qvar1}
\int \hskip -1mm d^4 \theta \, &  (Q_+ + Q_-) {\cal S}\\
  &= i \int \hskip -1mm d^4\theta \, \left( \bar\theta^+   \theta^-  \partial_+ + \bar\theta^-   \theta^+  \partial_-\right)  ( Q_+ + Q_-) {\cal S} - \int \hskip -1mm d^4\theta \, \theta^4   (Q_+ + Q_-) \partial_- \partial_+ {\cal S}\nonumber\\
&= i \partial_+ \bigr[  ( Q_+ + Q_-) {\cal S} \bigl]_{\theta^+ \bar \theta^-}  +\,  i \partial_- \bigr[  ( Q_+ + Q_-) {\cal S}\bigl]_{\theta^- \bar \theta^+}   - \partial_- \partial_+  \bigr[ (Q_+ + Q_-)  {\cal S}\bigl]_{\emptyset}.\nonumber
\end{align}
 Here $[ X ]_{\Theta}$ is the coefficient of the monomial $\Theta$ in the expansion of the superfield $X$, and $ [ X ]_{\emptyset}$ is its  bottom component. Multiplying by the anticommuting parameter
  $\epsilon$,  adding the complex conjugate and integrating over $M$ gives
 \bea\label{3.6} 
\int_{ M} \hskip -1mm d^2x  \hskip -1mm  \int \hskip -1mm  d^4\theta\,  {\cal D}_{\rm susy}  {\cal S}  =  \int_{\partial M}
 \hskip -1mm 
 dx^0\,     \left[ \frac{i}{2} \left(\bigr[{\cal D}_{\rm susy} {\cal S}\bigl]_{\theta^+ \bar \theta^-}  - \bigr[{\cal D}_{\rm susy} {\cal S}\bigl]_{\theta^- \bar \theta^+} \right)+ \frac{1}{4} \partial_1 \bigr[{\cal D}_{\rm susy} {\cal S}\bigl]_{\emptyset} \right]\ .  
\eea
Since $[{\cal D}_{\rm susy} X ]_{\Theta}$ is   the supersymmetry transformation of
 the superfield component $[  X ]_{\Theta}$, we have succeeded in what we set out to do,  which was to find a
 boundary term that compensates the supersymmetry transformation of the bulk D-term. 
 From \eqref{3.6} one  reads   
    \bea\label{3.7} 
 [{\cal S}]_{\rm bnry}  =  -    \frac{i}{2}  \left(  \bigr[  {\cal S}\bigl]_{\theta^+ \bar \theta^-} -  \bigr[ {\cal S}\bigl]_{\theta^- \bar \theta^+} \right)
  -  \frac{1}{4} \partial_1 \bigr[  {\cal S}\bigl]_{\emptyset} \ . 
 \eea 
It is of course also possible to verify the invariance of  \eqref{I} with the above $[{\cal S}]_{\rm bnry}$
by  using the transformations of a general superfield given in ref.\,\cite{Closset:2014pda}.

 The  supersymmetric completion \eqref{I}  depends  only on  the D-term density ${\cal S}$. It is not unique, 
 since  there  exist in general  many other combinations of boundary fields  that are 
  invariant   under the B-type supersymmetry. 
   After completing  the bulk Weyl  anomaly   
    in this generic `reference'  way,  we  can restrict our 
search for  extra  boundary contributions  
   to  terms that are separately super-invariant.

 
\subsection{Boundary superspace and invariants}

  A systematic method to  construct  boundary superinvariants is by using the formalism
of the boundary superspace  \cite{Hori:2000ic, mirror}. The 
   boundary superspace of B-type  is defined by the following  identifications of coordinates
\bea\label{bryspace}
x^+ = x^-\,, \qquad   \theta \equiv e^{-i\beta}\, \theta^+ =  e^{i\beta}\, \theta^-, \qquad 
\bar \theta \equiv e^{i\beta}\, \bar \theta^+ = e^{-i\beta}\,\bar \theta^-\ .  
\eea
These identifications are such that the generators \eqref{Dsusy} involve only a derivative 
in the time direction. 
We   set  again
  $\beta=0$. The supercovariant boundary derivatives  are
$$
\bar D = \bar D_+ + \bar D_- = - \frac{\partial}{\partial \bar\theta} + i \theta\, \partial_0\,,
 \qquad D = D_+ + D_- =  \frac{\partial}{\partial \theta} - i \bar\theta\, \partial_0\,.
$$
The restriction of a chiral bulk superfield  on a B-type boundary is chiral on the boundary, i.e.~it is annihilated by $\bar D$, whereas the  twisted-chiral bulk superfields  
   do  not have any  chirality property on the boundary. 
In particular
\bea\label{bsuperfield}
  \Sigma|_{\partial M} =  
   \sigma+ia  +\theta \bar \chi_+ + \bar \theta \chi_-   + \theta \bar\theta [w -i 
 \partial_1(\sigma+ia)]   
\eea
(with all  fields   functions of $x^0$)  is neither chiral nor antichiral, and the same is true for the coupling
superfields $\Lambda^I|_{\partial M}$. 
The usual  D-term and F-term integrals of   boundary superfields are  invariant under
B-type supersymmetry.   
\smallskip
 
    Let us consider as a special case the following composite bulk superfield: 
\bea\label{SF}
\delta {\cal S}^H  = -{1\over 4\pi}  ( \delta \Sigma +  \delta \bar\Sigma)\bigl[ H(\Lambda^I)  + \bar H(\bar \Lambda^{\bar I}) \bigr]
\eea
with $H$ a holomorphic function of the $\Lambda^I$.
 This  is the D-term integrand that entered  in  the K\"ahler-Weyl  transformation
of the  anomaly, eq.\,\eqref{KW1}. Recall that for a closed surface $M$,    $\int_M  d^2x \int d^4\theta\,  {\cal S}^H$  
is  a local gauge-invariant counterterm which can be 
 expressed as a twisted F-term 
in terms of the curvature superfield  $ \bar D_+ D_- \bar \Sigma$. 
The question we would like to ask  is whether
 this local   counterterm can be defined  when   $M$ is open. 
\smallskip

The  superspace  integral of $\delta {\cal S}^H$ is given by   eqs.\,\eqref{2} and \eqref{5} with 
 $K$ replaced by  $H + \bar H$. Adding the reference  completion \eqref{I}, \eqref{3.7} 
   gives after some  rearrangements  
\bea \label{KW4}
 {\rm I}_D({\cal S}^H)   
   &=& \hskip -2mm  {1\over 4\pi}  \int_M \hskip -0.7mm 
  d^2x \left[  (H+\bar H) \square \sigma  -i (H-\bar H)\square   a
 \right]   + {i\over 4\pi  }\int_{\partial M}  \hskip -0.7mm  dx^0 (  w \bar H -  \bar w H) \nonumber\\
   &+&\hskip -1mm  {1\over 4\pi}    \int_{\partial M}  \hskip -0.7mm  dx^0 \left[  \sigma\, \partial_1 (H+\bar H) + i \partial_1   a (H - \bar H)
   + {i\over 2} (  w +   \bar w) (H - \bar H)  \right] \ .   
\eea
The bulk integral   is the   local   counterterm that we encountered already
  in   section 2. The boundary integral, on the other hand,  involves the Weyl factor $\sigma$ 
  with no derivatives, so it cannot be written as the integral of a covariant density. 
  Fortunately  the   lower line of \eqref{KW4}
    is a  boundary superinvariant by itself (this is the reason why we isolated it). It is   the
    boundary D-term\  \,  $ \int  \hskip -0.7mm  dx^0 [{\cal B}^H]_{\theta\bar\theta}$ \ with 
   \bea\label{311}
   {\cal B}^H=  {i\over 8\pi} ( \Sigma +   \bar \Sigma) (H - \bar H)\vert_{\partial M}\ .   
 \eea 
 Subtracting this integral  from \eqref{KW4} gives then a  bulk and boundary counterterm, 
\bea\label{312}
   {\cal C}^H :=  \  {\rm I}_D({\cal S}^H) 
    -\int \hskip -1mm  dx^0 \,   [{\cal B}^H]_{\,\theta\bar\theta} \  , 
\eea
which   is   local,  supersymmetric  and gauge invariant.
It  can be used to compensate K\"ahler-Weyl  transformations
   when  $M$ is  open.  
     As a check, note that we could have written  ${\cal C}^H$ directly  
   as the more familiar boundary   completion of an F-term \cite{Hori:2000ck}. Indeed
\bea\label{IF}
    {\rm I}_F(\Phi) := \int_{ M} d^2x \hskip -0.7 mm  \int \hskip -1mm d \theta^+ d\bar \theta^-\, \Phi 
   \ - i \int_{ \partial M} dx^0\, [\Phi]_{\emptyset}
\eea
  is  type-B super-invariant for any twisted chiral field $\Phi$, 
  and\,  ${\cal C}^H = {\rm I}_F(\Phi) + c.c.$\  \,  for the 
  twisted chiral field 
  $\Phi =   - {1\over 4\pi}\,  \bar D_+D_-   \bar \Sigma\, H$. We leave the proof of these
  statements   as an exercise for the reader.

 
  \section{The  Boundary Anomaly}   

  We   have now the  tools at hand  to discuss the
   supersymmetric boundary  completion of the anomaly of   \cite{Gomis:2015yaa}.
   Recall that the bulk   anomaly 
   is given by the D-term integral 
 \bea
   A_{\rm closed} = \int_M d^2x\,   [ \delta {\cal S}]_{\rm top} \qquad {\rm where}\qquad 
 {\cal S} =  {1\over 4\pi } \left[ \,  {c\over 6} \Sigma \bar\Sigma  - (\Sigma +  \bar\Sigma) K  \right]\ . 
 \eea
As was explained in section 3, the most general boundary completion that is consistent with
 type-B supersymmetry can be written as 
 \bea\label{Aopen}
 A_{\rm open} =  \int_M d^2x\,   [ \delta {\cal S}]_{\rm top}  +  \int_{\partial M}
  \hskip -1mm  dx^0 \, ([\delta {\cal S}]_{\rm bnry} +  
  [\delta {\cal B}]_{\,\theta\bar\theta}) \ =\  {\rm I}_D(\delta {\cal S}) + \int_{\partial M}
  \hskip -1mm  dx^0 \,[\delta {\cal B}]_{\,\theta\bar\theta}
  \eea
 where $\delta {\cal B}$ is  a boundary superfield  made out of  $\delta \Sigma|_{\partial M}$, 
    $\Sigma|_{\partial M}$ and   $\Lambda^I|_{\partial M}$, and depending in general on the boundary condition
    $\Omega$.  
   Our task is now   to  find  $\delta{\cal B}$.
    
     Since we will need it later, let us record here the reference completion  $[\delta {\cal S}]_{\rm bnry} $\,. 
             The relevant terms
      in the superfield expansions are
$$
\Sigma \bar \Sigma = \sigma^2 + a^2 + \theta^+ \bar\theta^- w (\sigma - i a) + \theta^- \bar\theta^+ \bar w (\sigma + i a) + \ldots\ \ , 
$$ \vskip -9mm
\bea\label{41s}
(\Sigma +  \bar \Sigma) K = 2 \sigma K +  \theta^+ \bar\theta^- w K + \theta^- \bar\theta^+ \bar w K+ \ldots\ \ ,
\eea
where the dots stand for  other monomials in $\theta$,   or   terms that involve the fermions and
the  auxiliary fields in  $\Lambda^I$  which are set to zero.
 Plugging \eqref{41s} in  \eqref{3.7} 
gives  
 \bea\label{SBanomaly}
[ {\cal S}]_{\rm bnry} =  -  {1\over 8\pi}  \left[ \, {c\over 6}   
\Bigl(  \sigma \partial_1  \sigma + a \partial_1 a  +   a {(w+\bar w)} + i \sigma (w-\bar w)
    \Bigr) 
-   \bigl( \partial_1 (\sigma K)  + i K (w-\bar w)\bigr) 
  \right].
\eea \vskip 1mm
 \noindent  
 The contribution  to the anomaly, $\int_{\partial M} [\delta {\cal S}]_{\rm bnry} $,  is   the variation of 
 the above expression  with respect to the supergravity fields $\sigma, a$ and 
 $w$.

    
\subsection{Parity and  a general ansatz}

 An important restriction on  $\delta {\cal B}$  comes from  parity invariance.
   The  parity 
    transformation of B-type acts on  superspace 
    coordinates as follows~\cite{Brunner:2003zm}:
$$
x^+ \,\leftrightarrow\, x^-\, , \quad    e^{-i\beta}\, \theta^+ \,\leftrightarrow\,  e^{i\beta}\, \theta^-\, , \quad 
  e^{i\beta}\, \bar\theta^+ \,\leftrightarrow\,  e^{-i\beta}\, \bar\theta^-\, , \qquad {\rm and\ \ hence}
\qquad  y^\pm \,\leftrightarrow\,  \bar y^{\,\mp}\ . 
$$
  The  superspace boundary \eqref{bryspace} is the invariant locus of this transformation. 
  Parity conjugates  the Weyl  superfield,  $\Sigma \,\leftrightarrow\,  \bar\Sigma$, or explicitly
   in components 
\bea\label{parity1}
  \sigma \to \sigma\, , \quad a \to -a\, , \quad  e^{-i\beta}\,\chi_+  \,\leftrightarrow\,  e^{i\beta}\,\chi_-\, ,  \quad
w \to \bar w\  \ . 
\eea
That $a$ is indeed a pseudoscalar follows from its definition, 
  $V_\mu = \epsilon_{\mu\nu} \partial^\nu a$. 
We have introduced in  the above transformation  the  axial   phase that entered  in the preserved
supersymmetry.  
We  will continue to set   $\beta = 0$ in what follows. 

The  ${\cal N} = (2,2)$  SCFT is in general not parity invariant. Since
 B-type  parity conjugates the twisted chiral fields,  ${\cal O}^I \leftrightarrow \bar {\cal O}^{\bar I}$,  
the terms coupling to Im$(\lambda^I)$ are parity odd.  The symmetry  can be   restored 
if the 
    couplings also transform  like twisted chiral   fields,\footnote{Frequently in  the literature  
    the parity-odd terms 
      couple to the {\it real}  parts  of  the complex  moduli.
This  convention  and ours differ by  the redefinition  $\lambda^I \to i\lambda^I$.} 
   \bea\label{parity2}
  \Lambda^I \ \,\leftrightarrow\,   \   \bar \Lambda^{\bar I}\ .  
\eea
Invariance under \eqref{parity2} implies the  reality condition 
 $K(\lambda, \bar\lambda) = K(\bar\lambda,  \lambda)$ which follows also  from 
 the general form of  the instanton expansion of  $K$, see for instance \cite{Jockers:2012dk}. 
 
  The Zamolodchikov metric $\partial_I \bar\partial_{\bar J}K$ is   parity invariant, 
  while the  
 K\"ahler  one-form ${\cal K}$   is parity-odd. 
 It follows that   both  the bulk anomaly \,\eqref{1}\,-\,\eqref{5},  and its reference  completion 
   \eqref{SBanomaly},  are  parity-invariant  as should have been  expected.
      Note that the integrand  of  boundary terms must  change sign under parity
because of the change
  in the orientation of the surface.\footnote{For example  $\int d^2x\,   \partial^\mu \partial_\mu \phi
= \int dx^0\,  \partial_1\phi$\  is invariant if $\phi$ is a scalar, 
in which case $\partial_1\phi \to -\partial_1\phi$ under parity.
}
Note also that  invariance  under parity  rules out a  
bulk  anomaly  $ \sim \int_M \int d^4\theta\,  (\delta   \Sigma - \delta  \bar{  \Sigma})
L( \Lambda,\bar{ \Lambda})$  with $L$ a real  parity-even   function of the couplings, 
even though this   is  supersymmetric, 
cohomologically non-trivial and it obeys the Wess-Zumino consistency
 condition.
 \smallskip 
 
 Let us go back  now to the boundary anomaly  $\int 
  \hskip -1mm  dx^0 \,  
  [\delta {\cal B}]_{\,\theta\bar\theta} $, or better to its integrated form $\int 
  \hskip -1mm  dx^0 \,  
  [ {\cal B}]_{\,\theta\bar\theta} $  (locality and 
    Wess-Zumino consistency ensure that this latter form exists).
    The covariance and scale invariance of the  differential anomaly,   the fact that
    the central charge $c$ is constant on ${\cal M}$, and parity invariance severely constrain the allowed
    form of   ${\cal B}$ which is at most quadratic
 in the Weyl superfield $\Sigma\vert_{\partial M}$.
   The  general ansatz reads
  \bea\label{calB}
  {\cal B}   =  {i\over 8\pi} \left[  \#  {c\over 12} (   \Sigma^2    -   \bar  \Sigma^2)
  + \bar \Sigma\,  G^\Omega(\Lambda, \bar \Lambda) - \Sigma\,  G^\Omega(\bar \Lambda,  \Lambda)
  \right]  \Biggl\vert_{\partial M} 
  := {\cal B}^{(1)} + {\cal B}^{(2)}\ , 
  \eea
  where $G^\Omega$ is a function of the couplings (and of the boundary $\Omega$)
   obeying  the reality condition 
  \bea
  G^\Omega(\bar \Lambda,   \Lambda) =  [G^\Omega(\Lambda, \bar \Lambda)]^\star\ . 
  \eea
 The meaning of  $\vert_{\partial M}$ is that all  bulk superfields must be evaluated at the superspace boundary. 
 As in section 2,  we have separated  the central-charge anomaly
${\cal B}^{(1)}$   from  the moduli-dependent anomaly
${\cal B}^{(2)}$. 
        We have also extracted in the above expression certain numerical factors by anticipation. 
    They could be  reabsorbed in the definition of $G^\Omega$ and of  the (a priori arbitrary)  coefficient
       denoted  $\#$. We will now show  that $\# = 1$.

   In order to  fix this  coefficient  we  consider  pure Weyl transformations for which  
$\delta a = \delta w =   \delta\chi_\pm = 0$. Using the top component  of the boundary superfield 
 \bea \nonumber 
  [\delta {\cal B}^{(1)}]_{\theta\bar\theta} \ = \  {c\over 24\pi} 
    \bigl[ \, \sigma\partial_1\sigma  -  a\partial_1 a\  +{i\over 2}  \sigma\, (w-\bar w)  - {1\over 2} a\, (w+\bar w) -
 {i\over 2} (\bar\chi_+ \chi_- - \bar\chi_-\chi_+) \bigr] \ , 
 \eea
 together with eqs.\,\eqref{2}, \eqref{5} and  \eqref{SBanomaly},  
 leads to   the following central-charge   anomaly  on a surface $M$ with boundary:
 \bea\nonumber
 -i \delta_\sigma  \log {\cal Z}_V \supset 
  -{c\over 12 \pi}  \int \hskip -0.9mm d^2x \,   \delta\sigma \square \sigma \, + \,  {c\over 24 \pi}  \int 
 \hskip -0.9mm dx^0\, \bigl[ (1+ \# )
  \delta\sigma\, \partial_1\sigma
 - (1- \#  ) ( \sigma\, \partial_1 \delta\sigma 
 -  \delta\sigma\, {\rm Im} w)\, \bigr] \,. 
 \eea \vskip 1mm
\noindent This should be compared to the  standard  Weyl anomaly of a  CFT
  on  an open Riemann 
 surface.
Covariance and Wess-Zumino consistency fix completely the boundary anomaly, in this case
(see chapter 3  of  Polchinski's book  \cite{Polchinski:1998rq}) with the result
\bea\label{Polc} 
  A_{\rm book}= {c\over 24\pi } \, \Bigl[    \int_{ M}   \delta\sigma\, \sqrt{g} R +  2  \int_{ \partial M}  \delta\sigma\,  k\,  \Bigr]
=
    - {c\over 12\pi} \, \Bigl[  \int d^2x \,   \delta\sigma \square \sigma  -
    \int \hskip -0.8mm  dx^0 \,  \delta\sigma \partial_1  \sigma  \Bigr]  \ .    
\eea
Here $k$ is the extrinsic curvature of the boundary,  equal to  the  outward-normal derivative of the Weyl factor.\footnote{{In Minkowski signature and in  
conformal gauge  the general expression 
for the extrinsic curvature is   $k=t^\mu n_\nu \partial_\mu t^\nu - t^\nu t_\nu n^\mu \partial_\mu \sigma $, 
where $t^\nu$ and $n^\mu$ are the unit tangent and outward normal}  vectors. 
Here we work with a timelike boundary so $t^\nu t_\nu = -1$.
For the extension of the standard   Weyl anomaly to higher dimensional
manifolds with boundary see ref.\,\cite{Solodukhin:2015eca}\,.}
Matching  our result to \eqref{Polc}  shows  that   $\# =1$. 
Notice that  the dangerous term 
 $  \sigma \partial_1 \delta\sigma$,   which  would have given
  a non-covariant anomaly, 
  drops  out  
 for  $\# = 1$,  as   does the term depending on the auxiliary field $w$.


 \subsection{The holomorphic boundary charge}

 We turn next to the moduli-dependent term  ${\cal B}^{(2)}$ which features the 
 new  function $G^\Omega$. We will 
  show 
  that  this   must be of the  form
\bea\label{GOm}
  G^\Omega(\lambda, \bar\lambda) = K(\lambda, \bar\lambda) + 2 h^\Omega(\lambda)  \ , 
\eea
where  $h^\Omega$ is  a holomorphic function of $\lambda^I$. Parity invariance and reality 
of the anomaly 
 impose the  condition $\bigl(h^\Omega(\bar \lambda)\bigr)^\star 
= h^\Omega(\lambda)$, which means that this  holomorphic function
 admits  expansions with real coefficients.
 Later,  we will identify  $\exp(h^\Omega)$  with the  
central charge of  the boundary state. 

The starting point for proving \eqref{GOm} 
 is the linearized  coupling of the SCFT to the $U(1)_V$ supergravity fields \cite{Closset:2014pda, Gomis:2015yaa}. 
 Following  appendix C of ref.\,\cite{Gomis:2015yaa}, we  assume 
   that  in superconformal
 gauge
   this coupling can be written    as  a twisted F-term, 
 \bea
\int  \delta {\cal L}_{\rm sugra} = \ \int_M  \int d\theta^+ d \bar\theta^- (\delta \Sigma\, T)  - i   \int_{\partial M} [\delta\Sigma\, T]_{\emptyset} \ + 
 c.c.\ , 
 \eea
where $T$ is a twisted-chiral field
related to the ${\cal R}$-supermultiplet of the 
 energy-momentum tensor, i.e.~the supermultiplet  whose lowest component is the vector-like R-symmetry current. 
The boundary term in the above expression  is the one 
 that preserves  the supersymmetry of   twisted F-terms, eq.\,\eqref{IF}.  
If  we set  $ \delta\bar  \Sigma =0$,   this   coupling  is  $\bar Q_B$-exact, with 
  $\bar Q_B = \bar Q_+ + \bar Q_-$\,. This  fact follows from the simple  identity
  $$
     F(x) =  \{   {\bar Q}_B, [ Q_+, \phi(x) ] \} + i \partial_1 \phi(x)
  $$
  which is valid for  any twisted chiral field  with components $(\phi, \psi_-, \bar\psi_+, F)$. 
The same argument actually shows  that the  marginal deformations of the SCFT
 $$ 
 \int \delta {\cal L}_{\rm SCFT} =  {1\over \pi} \int_M \int d\theta^+ d \bar\theta^- (\Lambda^I \Phi_I)  - {i\over \pi}
    \int_{\partial M} [\Lambda^I\Phi_I]_{\emptyset} \ + 
 c.c.\ , 
$$
would be   $\bar Q_B$-exact if $\bar\Lambda^I = 0$\,. Here $\Phi_I$ is the twisted-chiral superfield whose top
component is the marginal operator ${\cal O}_I$. 
 Since $\bar Q_B^2 = 0$ and $\bar Q_B$ annihilates the
boundary state, we  conclude   that for $\delta\bar  \Sigma = \bar\Lambda^I =  0$ 
the following Ward identity holds
\bea\label{Ward}
\left\langle  \int  \delta {\cal L}_{\rm sugra}  \int \delta {\cal L}_{\rm SCFT} \right\rangle = 0\ . 
\eea
Put differently, there cannot exist terms  in the effective action 
proportional to $\delta   \Sigma\, \Lambda^I$  or to
$\delta  \bar  \Sigma\, \bar \Lambda^{\bar I}$. 
 Since all we  use  is  type-B supersymmetry which is preserved by the regulator, this    identity is   exact  even after including  contact terms. 

  Let us   first check that it   is verified 
 by   the bulk anomaly  \eqref{2}. The  relevant terms in $A^{(2)}$ have  derivatives acting  only on the
    holomorphic or  antiholomorphic couplings, but not both (the term  $\sim \partial\lambda \partial\bar\lambda$ gives  an
    anomaly of a mixed correlator to which the above  argument does not apply). 
       After  integration by parts of  $\int K \square \delta \sigma$ 
       one finds
   $$
  A^{(2)} \supset  -{1\over 4\pi}\int_M d^2x\, \Bigl[  \,\partial_\mu (\delta \sigma - i\delta a)\,\partial_I K\, \partial^\mu\lambda^I +
    \,\partial_\mu (\delta \sigma + i\delta a)\, \partial_{\bar I} K \, \partial^\mu\bar \lambda^{\bar I} \Bigr]\ . 
   $$
 This vanishes  if  $(\delta\sigma - i\delta a) = \bar\lambda^{\bar I} = 0$,
 as advertized. 

   Consider next  the  boundary anomaly. Its  general    form,   
eq.\,\eqref{Aopen},   includes  three different contributions: (i) the total derivatives 
   in \eqref{2} and \eqref{5} to which we should add the 
term   from  integrating  by parts  $\int K \square \delta \sigma$,  as just discussed; 
 (ii) the reference boundary completion 
\eqref{SBanomaly};  and (iii) the boundary superinvariant $\int_{\partial M} [{\cal B}]_{\theta\bar\theta}$
with ${\cal B}$ given by eq.\,\eqref{calB}. We focus  on the moduli-dependent anomaly. Collecting everything  gives
\bea\nonumber
 \,\hskip -9mm 
   - i  &&  \hskip -6mm
   \delta   \log {\cal Z}_V      \supset \  {1\over 4\pi} \int \hskip -0.9mm dx^0\,  
  \Biggl[  \partial_1 (\delta \sigma K)  + {i\over 2} (\delta w-  \delta \bar w) K \\
 && \,\hskip  -6mm
  +  {i\over 2} \left( (\delta\sigma -i\delta a) i (\bar\partial_{\bar I} G^\Omega\partial_1\bar\lambda^{\bar I} - \partial_IG^\Omega \partial_1\lambda^I) +
   (\delta\bar w + i \partial_1(\delta\sigma -i\delta a)) G^\Omega
  - c.c.  \right)  \Biggr]\ ,
  \label{413}
\eea
where the top line is the sum of contributions (i) and (ii), and the lower line is the contribution  (iii). 
This latter is the integral of 
 \bea\label{b2}
  [{\cal B}^{(2)}]_{\theta\bar\theta} \ = {i\over 8\pi} \left[ (\sigma -ia) i (\bar\partial_{\bar I} G^\Omega\partial_1\bar\lambda^{\bar I} - \partial_IG^\Omega \partial_1\lambda^I) +
   (\bar w + i \partial_1(\sigma -ia)) G^\Omega
  \right] + c.c. 
\eea      
 as follows from  the boundary  restrictions of $\Sigma$, eq.\,\eqref{bsuperfield},  and of $G^\Omega$, 
 $$G^\Omega(\Lambda, \bar\Lambda)\Bigl\vert_{\partial M}  = G^\Omega + i\theta\bar\theta\, 
 (\bar\partial_{\bar I} G^\Omega\partial_1\bar\lambda^{\bar I} - \partial_IG^\Omega \partial_1\lambda^I)\ . 
 $$        
 
    The expression \eqref{413} does not,  in general,  vanish 
    when $ \delta \bar \Sigma = \bar\lambda^{\bar I} = 0$. One notes however that
  for $G^\Omega =K$ the sum
    of the top and bottom lines collapses to
 $$
{1\over 4\pi} \int \hskip -0.9mm dx^0\,   (\delta\sigma  -i\delta a) \partial_I K 
  \partial_1\lambda^I + c.c. \ , 
 $$ 
which does have the desired property. 
Furthermore,  the lower line would vanish separately  if and only if 
    $G^\Omega$ were a holomorphic function of the couplings.\footnote{Recall that $\lambda^I$ are
    deformation parameters, and we can set $G^\Omega(0,0) =0$. }
 This establishes the general form of $G^\Omega$, eq.\,\eqref{GOm}. 

  We should also examine how  the anomaly  changes under K\"ahler-Weyl transformations.
  These latter act as follows on the bulk superfield  ${\cal S}$ and the boundary superfield ${\cal B}$
  that enter in the expression  \eqref{Aopen}: 
$${\cal S} \to  {\cal S}+ {\cal S}^H \ , \quad {\rm and}\quad 
{\cal B} \to {\cal B} + {i\over 8\pi} (\bar\Sigma - \Sigma) (H+ \bar H)\Bigr\vert_{\partial M}
+ {i\over 4\pi} (\bar\Sigma \,  \Delta  h^\Omega  - \Sigma \,  \Delta  \bar h^\Omega)\Bigr\vert_{\partial M}
$$
where $\Delta  h^\Omega$ denotes  the transformation of $h^\Omega$. 
Now  use $I_D(\delta{\cal S} + \delta{\cal S}^H) = I_D(\delta{\cal S}) + I_D(\delta{\cal S}^H)$, 
and the local counterterm,   ${\cal C}^H$,
that compensates  K\"ahler-Weyl  
 transformations, and  which we computed in eqs.\,\eqref{312}, \eqref{311}. 
Putting these two facts together implies  that the anomaly transforms  (in the sense of cohomology, i.e.~up to local counterterms)   as
$$
\Delta_{\rm KW} A_{\rm open}\  \simeq\  
 {i\over 4\pi}  \int dx^0  \left[ \delta \bar\Sigma (H + \Delta  h^\Omega) 
  - \delta \Sigma (\bar H + \Delta  \bar h^\Omega )  \right]_{\theta\bar\theta}\ . 
$$
Invariance   then implies that under K\"ahler-Weyl transformations
\bea\label{415h}
h^\Omega \to h^\Omega  - H\ ,
\eea
which means that  $e^{h^\Omega}$ is   a section of a holomorphic line bundle. 


\section{Anomalous one-point functions}

Taking stock of the analysis of the previous section, we can now write   the complete moduli-dependent
part of the super-Weyl anomaly,  
\bea
A_{\rm open} \supset  {1\over 4\pi}  && \hskip -7mm \int 
\hskip -0.9mm d^2x\,   \bigl[ K \square \delta \sigma -   
2 \delta\sigma \, 
 \partial_\mu \lambda^I \partial^\mu \bar\lambda^{\bar J} g_{I\bar J} +  2 (\partial_\mu \delta a)  {\cal K}^\mu
   \, \bigr]  +   {1\over 4\pi} \int \hskip -0.9mm 
 dx^0\,\Bigl[  -  K \partial_1\delta\sigma  \nonumber   \\
&&  
 +  (\delta\sigma  -i\delta a) \partial_I (K  + h^\Omega)
  \partial_1\lambda^I +  i \bigl( \delta \bar w   +i \partial_1(\delta\sigma -i \delta a) \bigr)\,   h^\Omega + c.c. \Bigr]\,. 
  \label{51}
\eea
The terms that survive  when $\delta\sigma = -\delta \log\mu$\,  is constant,  capture scale anomalies
of correlation functions in the SCFT.  The bulk term multiplying $\,  \partial_\mu\lambda^I  
\partial^\mu\bar\lambda^{\bar J}$  corresponds to
the anomalous two-point functions discussed in 
section 2. Likewise, the boundary  term proportional to $  \partial_1\lambda^I$ corresponds
to anomalous one-point functions of marginal operators on the half-plane. In this section we will
compute these one-point functions, and compare  them to known results about
Ramond-Ramond charges of D-branes \cite{Ooguri:1996ck}. Matching the two  will lead to the identification
of $\exp (h^\Omega)$ as the boundary charge.

 Translation invariance and 
  the scaling dimension  $\Delta = 2$ determine the one-point  functions   up to   unknown coefficients,
  \bea\label{scale1}
\langle {\cal O}_I(x)\rangle_\Omega \  =\   d^{\,\Omega}_I \,  {\cal R}{1 \over \vert x_1\vert^2}\ =  \ 
d^{\,\Omega}_I \, \, \partial_1^2\, [ \Theta(-x^1)  \log \vert x^1\mu\vert\,  ]\ .  
 \eea
 We  have here  introduced 
 a differential regularization   similar to that of the   two-point functions,
 eq.\,\eqref{2point}. The step function 
 is  $\Theta=1$ in  the half-space $x^1<0$, and $\Theta=0$ outside. 
Indeed, for any twice differentiable test function the integral of the right-hand-side is finite
 at $x^1=0$. 
      With the help of the   identity $\partial_1 \Theta (-x^1) = - \delta(x^1)$ 
 one derives    the following 
   scale dependence of the partition function 
\bea\label{56}
-i \mu{d\over \partial \mu} \log {\cal Z}_V  \,  \supset \, 
- \mu{d\over \partial \mu}\int {d^2x \over \pi}\, \lambda^I  \,    d^{\,\Omega}_I\, {\cal R} {1 \over \vert x^1 \vert^2}\ 
=\ -{1\over \pi}  \int  \hskip -1mm {dx^0  } \, d^{\,\Omega}_I\, \partial_1 \lambda^I
\ . 
\eea 
 It matches precisely with \eqref{51},  for $\delta\sigma = -\delta\log\mu$,  provided that
 \bea\label{57}
  d^{\,\Omega}_I  = {1\over 4} \partial_I (K  + h^\Omega)\ . 
 \eea  
 Thus the one-point functions of marginal operators must be completely determined by $K$ and $h^\Omega$. 
     Notice that the 
    $d_I^\Omega$  are  K\"ahler-Weyl invariant,   as expected for the  data of a SCFT.     
         
                 These  relations  are   reminiscent of those obeyed by
   the  RR  charges of D-branes in Calabi-Yau compactifications   \cite{Ooguri:1996ck}. 
    We will see  that this is no coincidence.   
 
  One more  remark is   in order here. Contrary to 
  what happened for the two-point functions, which only had  a  scale anomaly,
  the one-point functions also suffer from a global axial anomaly captured by the contributions that do not vanish
  for constant $\delta a$.  This anomaly reflects a contact term 
    in the two-point functions
$\langle (\partial_\mu j^\mu_A)\, {\cal O}_I \rangle_\Omega$
where $j_A^\mu$ is the axial R-symmetry current. 
One could compute this contact term directly   
along the  lines of  refs. \cite{Osborn:1993cr, Erdmenger:1996yc}. 
Here we have derived it from the  supersymmetric Ward identity \eqref{Ward},  which  related it to the  scale  
 anomaly as  we have  explained.


 \subsection{Ramond-Ramond  charges}
    
         The RR charges of a D-brane are the overlaps of the corresponding boundary state 
         $\vert \Omega \rangle  \hskip -0.6mm \rangle$ with the supersymmetric vacua of the worldsheet
         theory. These  latter 
         are  of two kinds:  (i) the canonical ground state $\vert 0  \rangle_{\rm RR}$ obtained from the Neveu-Schwarz      
          vacuum  by spectral flow, and (ii)
             the states $\vert I  \rangle_{\rm RR}$  obtained by spectral flow from the  Neveu-Schwarz
     states
     $\phi^I(0) \vert 0 \rangle_{\rm NS}$,   where  $\phi^I$ are  the   lowest  components of  
     twisted-chiral superfields whose highest components are the marginal operators ${\cal O}^I$. 
         The  geometry of this  `vacuum bundle', i.e.~how the collection of these vacuum states varies as a function 
         of the moduli, has been described in the classical work of  Cecotti and Vafa 
          \cite{Cecotti:1991me, Bershadsky:1993cx}.       
             
             The boundary state is a formal sum of  Ishibashi states, one for each representation
              of the ${\cal N}= (2,2)$ superconformal
             algebra. The coefficients in this sum are determined by the
              inner product of $\vert \Omega \rangle  \hskip -0.6mm \rangle$ with the highest-weight states in each Ishibashi state. Of  particular interest is  the
             projection $\Pi_{\rm vac}$ of $\vert \Omega \rangle  \hskip -0.5mm \rangle$ on the supersymmetric
             ground states\footnote{A B-type brane has no component along ground states obtained 
             by spectral flow from chiral $(c,c)$ fields. So ground states here refers implicitly to those obtained
             by flowing from the twisted chiral $(a,c)$ fields only.}   
   $$
    \Pi_{\rm vac}\,  \vert \Omega \rangle  \hskip -0.5mm \rangle   := \, 
       c^\Omega\,   \vert 0  \rangle_{\rm RR} \ +\  \sum_I   c_I^\Omega \,  \vert I \rangle_{\rm RR}\  . 
     $$   
              The key observation of Ooguri, Oz and Yin  \cite{Ooguri:1996ck}  (see also \cite{Hori:2000ck})   is that 
         this   is a flat section of an improved connection $\nabla  = D  - C $ on the 
          (twisted-chiral)   moduli space, where
      $(C_I)^{J}_{\ K}$ are  the structure constants of the twisted-chiral ring, and  the overall
    normalization is chosen so that 
    $\vert 0  \rangle_{\rm RR}$ has   holomorphic dependence on the moduli. 
     In practice, for our purposes here,  these statements imply
     $$
         \bar\partial_{\bar I} \, c^\Omega = 0\ , \qquad 
          \partial_I c^\Omega
           +   (\partial_I K) \,
            c^\Omega -  c_I^\Omega = 0 \ ,  
     $$
   from which one finds easily
     \begin{equation}\label{eq:OOY_relation}
     \frac{c_I^\Omega }{c^\Omega } = \partial_I ( K + \log c^\Omega ).
     \end{equation}
We will now argue  that    $ {c_I^\Omega/ c^\Omega } = 4 d^{\,\Omega}_I $, so that  comparing 
  \eqref{57} with  \eqref{eq:OOY_relation}  identifies   $h^\Omega$ with the
logarithm of the canonical RR charge of the $\Omega$ brane, 
 \bea\label{eq:Cc0}
 h^\Omega(\lambda)  = \log c^\Omega(\lambda)\ . 
 \eea
     
     To   relate the one-point function coefficients    with the   RR charges
     of the boundary D-brane
    we  perform   the conformal map from  the half-plane to the semi-infinite cylinder, 
    This reads
      $$
        y  = {z+1 \over z-1} \  \quad \Longrightarrow\quad   {\partial y   \over \partial z} = - {2 \over (z-1)^2}\ , 
          $$
    where $z$ parametrizes the half-plane (Re$z \leq 0$ with $z = x^1 - i x^2$) and  $y$ parametrizes the
    cylinder ($\log y =  {\tau + i\varphi}$ with $\tau\leq 0$). 
   For any 
    conformal scalar primary field  $\Psi_{\Delta}$ with scaling dimension  $\Delta$ one has 
       $$
        \langle  \Psi_{\Delta}(z=  -1) \rangle_{\rm half-plane}  \ =  \    \lim_{\tau\to -\infty}\, {1\over 2^\Delta } \, e^{-\Delta \tau}
         \langle  \Psi_{\Delta} (y) \rangle_{\rm half-cylinder} \ =\ 
         {1\over 2^\Delta }  {\langle \hskip -0.6mm \langle \Omega \vert \Psi_{\Delta}  \rangle_{\rm NS}
          \over  \langle
         \hskip -0.6mm \langle \Omega \vert 0  \rangle_{\rm NS} }\ 
        ,  
       $$                               
where $\vert \Psi_{\Delta} \rangle$ is the state created by acting with  $\Psi_{\Delta}$  on the NS vacuum,
and the overall normalization was fixed so that insertion of the identity operator gives 1. 
     Pick   $\Psi_{\Delta} = {\cal O}_I$, so that $\Delta = 2$,  and insert in the inner products on the
     right-hand side 
      the spectral flow operator $e^{i \hat \xi}$. This latter maps NS states to RR states,\footnote{Usually the starting
       point of the spectral flow is the state created by the lowest component of the marginal superfield, but this is charged 
       so its  OPE with  $e^{i \hat \xi}$ is singular. Since we are interested here in  normalizations,  it is  preferable to
       start with the   top component which is neutral and has therefore a non-singular OPE. 
       As the end states of the flow   lie  in the same  Ishibashi block,  they have  the same
       coupling to the boundary state. 
      } 
      while its
    action  on the boundary state is a pure phase,
      $e^{i \hat \xi} \vert\Omega\rangle \hskip -0.5mm \rangle = e^{i\xi^\Omega} \vert\Omega\rangle \hskip -0.5mm \rangle $.  It follows easily that 
   \bea
   d_I^\Omega = {c_I^\Omega \over 4 c^\Omega } 
      \eea   
  which is the sought-for relationship.   
  
   Equations  \eqref{57} and   \eqref{eq:OOY_relation}
  have been  obtained from   different routes,  so the fact that they   match   is a confirmation of
   our  result for  the super-Weyl anomaly.


   \section{Hemisphere partition functions}        
 
    Up to now we have considered the generating functional of correlation functions  expanded
    around the flat-metric background, i.e.~for  $\delta\Sigma$ very   small.  
       In this section  we will  integrate the anomaly and  calculate ${\cal Z}(D^2)$
     for   supersymmetric   backgrounds with the topology of the disk. 
    We would like, in particular,  to
      prove  the conjecture   of refs.\,\cite{Hori:2013ika,Honda:2013uca}  that  the round-hemisphere partition
      function   computes the
    holomorphic boundary charge. 
 
         We begin by restricting the anomaly to the case of 
 constant   $\lambda^I$. Dropping all derivative terms  in 
  eq.\,\eqref{51}  (as well as the moduli-independent terms) 
 we find
\bea\label{61}
A_{\rm open}  \supset  \delta\, \Bigr\{
  -    {1\over 4\pi} \int \hskip -1mm {d^2x}\, 
  \Bigl[ \square ( \sigma - i  a) h^\Omega  + \square ( \sigma +  i   a) \bar h^\Omega 
   \Bigr]  
   +  {i\over 4\pi} \int \hskip -1mm 
   dx^0 \, \Bigl[   \bar w \, h^\Omega  -  w\,  \bar h^\Omega  \Bigr] 
   \Bigr\}
   \,. 
\eea    
As was the case for closed $M$, here also the constant-$\lambda$
  anomaly is the variation of a local covariant action.
It would have   been  cohomologically trivial   in a bosonic theory, but  acquires  universal
meaning thanks to  ${\cal N}=(2,2)$
 supersymmetry. 
\smallskip  
                                            
     The expression inside the curly brackets is the { integrated}  anomaly,   $-i \log {\cal Z}$.                                                                                   
  In writing it we have converted  the  boundary terms to  bulk integrals of total derivatives. 
   This does not change  the anomaly, but it ensures that 
   $-i \log {\cal Z}(M) \to 0$  when  $M$  is a  vanishingly-small  disk.                                                                                                                             Consider for instance   terms  in \eqref{51} that  are proportional to the
K\"ahler potential. If we convert  the boundary term to a total-derivative the  two such  terms
cancel each other. Otherwise
 $  K  (\int d^2x\, \square \sigma  - \int dx^0 \, \partial_1\sigma) = -2\pi K\, \chi_M$
with $\chi_M$                                                                                                                                                                                                                                                   the Euler characteristic of the surface.  Neither choice  contributes  to   $\delta \log {\cal Z}$,
but only the first one guarantees that  excising   a tiny bit of surface from $M$  (e.g.~as part of the regularization)
 does not change the free energy by a  finite amount. Furthermore,
  by converting the Wilson line to integrated flux we 
   make it  insensitive to gauge-choice singularities. The contribution of auxiliary fields 
   was left as a boundary integral with the implicit understanding
  that   $w$ is non-singular, i.e.~that $\oint _{\cal C} w \to 0$
   for any shrinking cycle ${\cal C}$ in the interior of $M$.


 \subsection{Killing spinor equations}
   
  In the integrated anomaly  \eqref{61} the dependence on the K\"ahler potential  dropped out,
  so ${\cal Z}(M)$  depends only on   the boundary charge. 
 To compute ${\cal Z}(M)$  we need, in addition to the metric and gauge field, also the  
 auxiliary fields $w, \bar w$. For   supersymmetric backgrounds
 these follow   from the Killing-spinor equations of  the ${\cal N}=2$   supergravity, 
  whose covariant form can be found    in  ref.\,\cite{Closset:2014pda}. 
  We take here the  simpler route \cite{Gomis:2015yaa} of  working   directly in
   superconformal gauge,  where these equations  reduce to the condition that the twisted-chiral-field
    background $\exp(\Sigma_{\rm backgr})$
  be annihilated by global  superconformal transformations.

       The standard supersymmetry  transformations of a twisted chiral field with components
            $(\phi, \psi_-, \bar \psi_+, F)$, and of the conjugate antichiral field,  
    read  
\begin{align*}
&\delta_{\rm susy} \phi = \bar \epsilon_+ \psi_- - \epsilon_- \bar\psi_+, & &\delta_{\rm susy} \bar\phi = \bar \epsilon_- \psi_+ - \epsilon_+ \bar\psi_-, \\
&\delta_{\rm susy} \psi_- = -2 i \epsilon_+ \partial_- \phi + \epsilon_- F, &&\delta_{\rm susy} \bar\psi_- = 2i \bar\epsilon_+ \partial_- \bar\phi + \bar\epsilon_- \bar F,\\
&\delta_{\rm susy} \bar\psi_+ = 2 i \bar\epsilon_- \partial_+ \phi + \bar\epsilon_+ F,
&&\delta_{\rm susy}\psi_+ = -2i \epsilon_- \partial_+ \bar \phi + \epsilon_+ \bar F, \\
&\delta_{\rm susy} F = -2i\epsilon_+ \partial_- \bar\psi_+ - 2i\bar\epsilon_- \partial_+\psi_-, &&\delta_{\rm susy} \bar F = -2i \bar\epsilon_+ \partial_- \psi_+ - 2i \epsilon_- \partial_+ \bar\psi_-.
\end{align*}                                             
Assume that $\phi$ is a conformal primary field with conformal dimensions $(\Delta_+, \Delta_-)$,  
 so that  the  fermions $\psi_+$ and $\psi_-$ have dimensions  $(\Delta_+ + {1\over 2}, \Delta_-)$
 and $(\Delta_+ , \Delta_- + {1\over 2})$ while the auxiliary field $F$ has dimensions 
 $(\Delta_+ + {1\over 2}, \Delta_-+ {1\over 2})$. 
   We may render the above equations covariant under arbitrary conformal transformations  by  letting
   the   parameters $\epsilon_-$ and  $\epsilon_+$ transform as conformal
  tensors of dimensions  $(-{1\over 2}, 0)$ and $(0, -{1\over 2})$, and by modifying  appropriately the  
  derivatives, 
$$
 \delta_{\rm susy} \phi = \bar \epsilon_+ \psi_- - \epsilon_- \bar\psi_+, \, \qquad
  \delta_{\rm susy} F = -2i\epsilon_+ D_- \bar\psi_+ - 2i\bar\epsilon_- D_+\psi_-,
$$ \vskip -9mm
 \bea\label{62}
  \delta_{\rm susy} \psi_- = -2 i \epsilon_+ D_- \phi + \epsilon_- F, \qquad
  \delta_{\rm susy} \bar\psi_+ = 2 i \bar\epsilon_- D_+ \phi + \bar\epsilon_+ F \ , 
 \eea
where  for a conformal tensor  $X$ with dimensions $(\Delta_+, \Delta_-)$   the covariant derivatives are
\bea\label{63}
\epsilon_+ D_- X\,  := \, \epsilon_+ \partial_- X +  2 \Delta_-  (\partial_- \epsilon_+) X\ , \qquad
\bar\epsilon_- D_+ X :=  \bar\epsilon_- \partial_+ X+  2 \Delta_+  (\partial_+\bar\epsilon_- )X \ . 
\eea
 Similar formulae apply to the
 twisted anti-chiral field. 
  The  modified transformations reduce to the standard ones for constant $\epsilon_\pm$,
  $\bar\epsilon_\pm$, and  behave homogeneously under changes of the conformal frame
 provided that
   $\epsilon_+$, $\bar\epsilon_+$ are   functions of  $x^-$,  and
    $\epsilon_-$, $\bar\epsilon_-$   functions of  $x^+$.
     Changes of frame    make these functions arbitrary. 
     
       Following  \cite{Gomis:2015yaa} we can express the Killing spinor equations as the conditions
       that   $ e^\Sigma $ be left invariant by  a  globally-defined superconformal transformation. 
       The exponential of the Weyl superfield is a
 twisted chiral  field with dimensions $({1\over 2},{1\over 2})$.\footnote{This  field  
  transforms homogeneously 
   under  the reparametrizations  and  $U(1)_V$ 
 gauge transformations that preserve  the  superconformal gauge: 
   $ 
 x^{\pm\, \prime}  =  f^\pm(x^\pm)$ and $V^\prime =  V + dg $ with $g = g^+(x^+)  +  g^-(x^-)\ . 
 $ 
A simple calculation gives
$
 \exp(\sigma^\prime  + ia^\prime )  =  \left( {df^+/dx^+}   \right)^{- {1/2}} \left( {df^-/ dx^-}  
  \right)^{-{1/ 2}}
 \exp(i g^+ - ig^-)   \exp(\sigma + ia) \ . 
$
Note that a $U(1)_V$ gauge transformation with angle 
 $(g^+ + g^-)$ is equivalent to a $U(1)_A$ gauge transformation with angle $(g^+ - g^-)$. 
 Thus the lowest component of $\exp(\Sigma)$ 
behaves under  this restricted class of   transformations 
as a  $({1/2}, {1/ 2} )$ conformal tensor  with unit   { axial-}R  charge.
  }        
Its bosonic components are $\phi = \exp(\sigma + ia)$,    $F =    \exp(\sigma + ia) w$, 
     $ \bar\phi = \exp(\sigma -  ia)$,    $\bar F   = \exp(\sigma - ia) \bar w$, 
while the   fermionic background is set to zero.  Invariance under  \eqref{62} then  implies  
  \begin{align}
 \epsilon_- w =& \hskip 3mm 2i  {\epsilon_+} \partial_- (\sigma + ia + \log \epsilon_+) \, , \qquad
  {\bar \epsilon_+}  w
 =  -2i  {\bar \epsilon_-}\partial_{+} (\sigma + i a +\log \bar\epsilon_-)\, ,
\nonumber \\
 {\epsilon_+}  \bar w =& \hskip 3mm
 2i 
 {\epsilon_-}\partial_{ +} (\sigma - ia + \log \epsilon_-)\, , \qquad
 {\bar \epsilon_-}  \bar w =   - 2i  {\bar \epsilon_+} \partial_- (\sigma - i a +\log \bar\epsilon_+)\, .
\label{weqs}
\end{align}
We are interested in   backgrounds with Euclidean signature, so we perform the
 Wick rotation  $x^0 \to -ix^2$ which maps  $(x^+, x^-)$ to $(\bar z, -z)$  
with  $z:= x^1+ix^2$.  We also write the Wick rotated $\epsilon$'s 
(which we continue to label  with subscripts `$\pm$') as constant anticommuting parameters 
 multiplying   holomorphic
or antiholomorphic Killing spinors 
\bea
\epsilon_+   =   \epsilon\,  \zeta^- (z) \ ,  \quad  \epsilon_-   =   - \epsilon\,  \zeta^+ (\bar z)\ , \quad
   \bar \epsilon_+  =    \bar \epsilon\,  \bar\zeta^- (z)\ , \quad 
    \bar \epsilon_-  =    - \bar \epsilon\,  \bar\zeta^+ (z)\ . 
\eea
Depending on the  unbroken supersymmetries 
some of these parameters could  be set  to zero. 
We are interested in supersymmetries  compatible with the B-type boundary conditions, for which
all the $\zeta$'s are non-vanishing.  
 With the above  conventions the auxiliary fields read
\begin{align} \nonumber
w =& \hskip 5mm 2i \frac{\zeta^-}{\zeta^+} \partial_z (\sigma + ia + \log \zeta^-) = \hskip 3mm 2i \frac{\bar \zeta^+}{\bar \zeta^-} \partial_{\bar z} (\sigma + i a +\log \bar\zeta^+)\, ,\\
\bar w =&  -2i \frac{\zeta^+}{\zeta^-} \partial_{\bar z} (\sigma - ia + \log \zeta^+) = - 2i \frac{\bar \zeta^-}{\bar \zeta^+} \partial_z (\sigma - i a +\log \bar\zeta^- )\, .
\label{ws}
\end{align} 
 These are the  relations derived in   appendix D of ref.\,\cite{Gomis:2015yaa}. 
 

 \subsection{The (squashed) hemisphere}                                                                                                                                  
                                                                                                                                                                                                                                                                   
To find non-trivial solutions of these equations one must allow for non-hermitean backgrounds 
in which the metric factor $\sigma$ is real,  but $a$ is allowed to
 be  complex and $(\bar w)^* \not= w$.  Consider  a surface  with  disk  topology  
 parametrized by    $\{ z\in \mathbb{C} ; \  \vert z\vert \leq 1\}$.   
 We are interested in solutions that obey the  B-type boundary conditions at $\vert z\vert = 1$:
 \bea\label{bnry6}
  \bar z^{\,-{1\over 2}}\, \zeta^+    =    e^{-2i\beta}\, z^{\,-{1\over 2}}\zeta^- \ , \qquad
  \bar z^{\,-{1\over 2}} \bar \zeta^+    =    e^{2i\beta}\, z^{\,-{1\over 2}}\bar \zeta^-\ .
 \eea
  Here $\beta$ is the axial phase introduced in section 3,  and we have transformed 
  to the cylindrical coordinate  $\log z = \tau + i\varphi$ 
  using  the fact that $\epsilon_+$ and $\epsilon_-$ are conformal tensors with weight
  $(0, -{1\over 2})$ and $(-{1\over 2}, 0)$. These  boundary conditions  admit  two inequivalent
  solutions for given  $\beta$. If  $\beta = 0$ they read
    \begin{align}   \nonumber
  \label{eq:spinors}
(+) :& \quad \zeta^- = 1, \quad \zeta^+ = \bar z, \hskip 6.5mm \, \bar\zeta^- = z, \quad \bar \zeta^+ = 1,\\
(-):& \quad \zeta^- = z, \quad \zeta^+ = 1, \qquad \bar\zeta^- = 1, \quad \bar \zeta^+ = \bar z\ . 
\end{align}
These  two choices,  related by CPT,  correspond to  the two
choices of spin structure in the conformal Killing spinor equation
 on the hemipshere~\cite{Hori:2013ika}.\footnote{In  \cite{Hori:2013ika}  the minus spin structure is
    defined with phase $\beta=\pi/2$,  which arises naturally from CPT.  Since for a single boundary the
    gluing phase is irrelevant, we   here set it to zero.
    } 
Inserting these expressions in  \eqref{ws} shows that both $\sigma$ and $a$ must be
 independent of the phase
of $z$. Indeed,  the solutions of the Killing spinor equations with the boundary conditions
\eqref{bnry6} leave unbroken a  $SU(1\vert 1)$ subgroup 
of the   global superconformal group $OSp(2\vert 2, \mathbb{C})$.  
 The Killing isometry corresponds to the $U(1)$ factor of this unbroken symmetry.

\smallskip

  A particular   supersymmetric background is the perfectly 
  round hemisphere with vanishing  $U(1)_V$ gauge-field:  
  $$
  \sigma =  -\log  (1 + z\bar z)  + {\rm constant} \ , \qquad a= 0\  . 
  $$
For this background  one finds the auxiliary fields
  $$
(+): \ \ w = \bar w = -\frac{2i}{1+z\bar z}\ , \qquad
(-): \ \ w = \bar w =  \frac{2i}{1+z\bar z}\ , 
$$
which  are  smooth in the interior and take constant values
on  the boundary,  $w = \bar w = \mp i$. 
 Inserting  the above expressions   in  \eqref{61},   and recalling  
that $h^\Omega = \log c^\Omega$,  leads finally  to the 
hemisphere partition functions
\bea\label{fnl}
Z_+(D^2, \Omega)  =  {\cal Z}_0  \, c^{\Omega}(\lambda)\, , \qquad \qquad
Z_-(D^2, \Omega) =   {\cal Z}_0 \,  c^{\Omega} (\bar\lambda) \, . 
\eea
This establishes the conjecture of  refs.\,\cite{Honda:2013uca},\cite{Hori:2013ika}. The 
moduli-independent factor  ${\cal Z}_0$\,\footnote{In principle, the free energy could contain non-local  terms
  that are Weyl-invariant and $a$-independent  
  and hence  make no   contribution  to the anomaly. We are  assuming that if such terms exist they
  are  $\lambda$-independent.}\,
  is a priori scheme-dependent,  and hence uninteresting. 
 It can be determined in accordance  with the 2-sphere  as in the eqs.\,\eqref{02}
 of the introduction.

      It is in fact straightforward to extend the calculation to more general   metric and 
   gauge-field backgrounds that respect the symmetry  under phase rotations of $z$.                
       An example is the squashed-hemisphere background of   \cite{Honda:2013uca, Gomis:2012wy}.                           
  Let $\delta\sigma(z\bar z)$ and $a(z\bar z)$ be the deformations of the round hemisphere background.
  Inserting   eqs.\,\eqref{ws} and \eqref{bnry6}  in the integrated anomaly (the expression
 in  curly brackets in \eqref{61}) shows that all  dependence on $\delta\sigma$ and $a$ drops out 
 as long as  the $z\to 0$ region is smooth, i.e.~free from conical and  Dirac-string singularities. 
 For all such  backgrounds  the results \eqref{fnl} continue  to hold.  
 
 \smallskip
 
      We conclude with some  remarks. First, it follows  from  the  K\"ahler-Weyl transformation
      \eqref{415h} that the hemisphere partition functions are  sections of  (anti)holomorphic line bundles.
      The    unambiguous quantities are the partition-function ratios  $Z_\pm(D^2, \Omega_1)/ Z_\pm(D^2, \Omega_2)$ 
      for pairs of different boundary conditions, as well as the 
      $g$-function \eqref{03} which is  the physical
      degeneracy of the boundary.  It is interesting that for  ${\cal  N}=2$ 
      boundaries the $g$-function can be determined  entirely by anomalies. 
 
        A second remark concerns the superconformal   interfaces that transport the SCFT along  its 
        moduli space  ${\cal M}$. These can be mapped to boundaries by  folding the surface along the interface
        and complex conjugating the folded theory \cite{Bachas:2001vj}. The 
         central charge and entropy 
        of such interfaces  depends on  the analytic extension of the K\"ahler potential, 
            \cite{Bachas:2013nxa}
        $$
        c^\Omega =  e^{-{1\over 2} K(\lambda_1, \bar \lambda_2)}\ , \qquad   2 \log  g^\Omega
        =   K(\lambda_1, \bar \lambda_1)  + K(\lambda_2, \bar \lambda_2) - 
        K(\lambda_1, \bar \lambda_2)  - K(\lambda_2, \bar \lambda_1)\ . 
         $$ 
  This  extension can be therefore computed by localization of the hemisphere partition function. 
   
       Finally,  it should be possible to extend the analysis of  
        \cite{Gomis:2015yaa} 
        to four-dimensional manifolds with boundary. 
       A localization calculation of  an ${\cal N}=2$ supersymmetric gauge theory on the four-dimensional hemisphere has been
       performed recently in  \cite{Gava:2016oep}. Another interesting question
       concerns the dependence of $c^\Omega$ on the boundary (or open-string)
       moduli. Since this is part of the definition of $\Omega$ it should be also
        accessible by  localization techniques. 
  
  \medskip
  \medskip
  \medskip
  
 \noindent {\bf \large Acknowledgements}
 
 \medskip
 
We are grateful to Jaume  Gomis for many enlightening conversations, 
and especially for pointing out 
 a problem with 
 the integrated anomaly   of  an earlier draft. 
We also thank   Benjamin Assel, Ilka Brunner, 
 Amir Kashani-Poor, Zohar Komargodski, Anatoly Konechny,  Tasos Petkou,
Adam Schwimmer and Stefan Theisen for discussions.  
The work of D.P. is  supported by  the grant ANR-13-BS05-0001.
 
 \vskip 1cm
 

\appendix
\section{Notation and conventions}
 
We follow the conventions in chapter 12 of \cite{mirror}. 
Superfields are functions on ${\cal N}=(2,2)$ superspace with coordinates $(x^\pm, \theta^\pm, \bar\theta^\pm)$, 
where $x^\pm = x^0\pm x^1$. The flat Minkowski 
metric is $\eta_{00} =  - \eta_{11}= -1$,  so that $\square = -4\partial_+\partial_-$.
The Wick rotation sets  $x^0 = -ix^2$. 
Complex conjugation flips the order of the fermionic coordinates and 
acts on them as    $(\theta^\pm)^* = \bar\theta^\pm$.
The  Grassmann  integration measure is $
d^4\theta := d\theta^+d\theta^-d\bar\theta^-d\bar\theta^+
$. 

A general   supersymmetry transformation  reads
\bea\label{A1}
\Delta_{susy} = \epsilon_+ Q_- - \epsilon_- Q_+ - \bar \epsilon_+ \bar Q_- + \bar\epsilon_- \bar Q_+
\eea
where  the $\epsilon$ are anticommuting parameters and
\bea
{\cal Q}_\pm = {\partial \over \partial \theta^\pm} + i \bar\theta^\pm \partial_\pm\ ,
\qquad
\overline{\cal Q}_\pm = - {\partial \over \partial \bar \theta^\pm} - i  \theta^\pm \partial_\pm\ .
\eea
  Another useful set of differential operators   is
$$
D_\pm = {\partial \over \partial \theta^\pm} - i \bar\theta^\pm \partial_\pm\ ,
\qquad
\overline{D}_\pm = - {\partial \over \partial \bar \theta^\pm} + i  \theta^\pm \partial_\pm\ .
$$
Twisted chiral fields obey the relations
$$
\overline{D}_+ \Phi = D_- \Phi = 0\ , 
$$
and have the following  expansion in components 
\bea\label{apptc}
\Phi = \phi(y^\pm) + \theta^+ \bar \psi_+(y^\pm) + \bar\theta^- \ \psi_-(y^\pm) + \theta^+\bar\theta^- F(y^\pm)\ , 
\eea
where $y^\pm = x^\pm \mp  i\theta^\pm \bar\theta^\pm$. 
The operators $D_\pm$ have trivial cohomology, meaning that $D_+{\cal F}= 0$ implies that
${\cal F}= D_+ {\cal G}$\,  for some superfield ${\cal G}$. 

The supersymmetry variation of a twisted chiral field in terms of components reads
 \begin{align*}
&\delta_{\rm susy} \phi = \bar \epsilon_+ \psi_- - \epsilon_- \bar\psi_+, & &\delta_{\rm susy} \bar\phi = \bar \epsilon_- \psi_+ - \epsilon_+ \bar\psi_-, \\
&\delta_{\rm susy} \psi_- = -2 i \epsilon_+ \partial_- \phi + \epsilon_- F, &&\delta_{\rm susy} \bar\psi_- = 2i \bar\epsilon_+ \partial_- \bar\phi + \bar\epsilon_- \bar F,\\
&\delta_{\rm susy} \bar\psi_+ = 2 i \bar\epsilon_- \partial_+ \phi + \bar\epsilon_+ F,
&&\delta_{\rm susy}\psi_+ = -2i \epsilon_- \partial_+ \bar \phi + \epsilon_+ \bar F, \\
&\delta_{\rm susy} F = -2i\epsilon_+ \partial_- \bar\psi_+ - 2i\bar\epsilon_- \partial_+\psi_-, &&\delta_{\rm susy} \bar F = -2i \bar\epsilon_+ \partial_- \psi_+ - 2i \epsilon_- \partial_+ \bar\psi_-.
\end{align*}

\smallskip

On a B-type boundary we have the following  identifications of coordinates, 
\bea
x^+ = x^- \ , \quad \theta := e^{-i\beta}\theta^+ =e^{i\beta} \theta^- \, , \qquad
\bar\theta := e^{i\beta} \bar\theta^+ = e^{-i\beta} \bar\theta^-\, . 
\eea
The unbroken supersymmetries are generated by \eqref{A1} with
  $\epsilon := e^{-i\beta}\epsilon_+ =  - e^{i\beta} \epsilon_-$\, 
and \, $\bar \epsilon := e^{i\beta} \bar \epsilon_+ =  - e^{-i\beta}\bar \epsilon_-$\,. Unless indicated otherwise, we set the phase $\beta = 0$.

The Euler density is  $\sqrt{g} R = -2 \square \sigma $  and the Euler characteristic reads
\bea
\chi_{\hskip -1.3mm \ _M} = 
{1\over 4\pi} \left[ \int_M \sqrt{g} R  + 2 \int_{\partial M} k\right] 
= -{1\over 2\pi} \left[ \int_M \square \sigma -  \int_{\partial M} \partial_\perp \sigma\right] 
= 2 -2h_{\hskip -1.3mm \ _M} -  b_{\hskip -1.3mm \ _M}
\eea
where $h_M$ is the number of handles  and $b_M$ the number of boundaries of the surface $M$.
 The normal derivative of the Weyl factor  is in the outward direction.


%

\end{document}